\documentclass{lmcs} 
\pdfoutput=1
\usepackage[utf8]{inputenc}

\usepackage{lastpage}
\lmcsdoi{19}{3}{5}
\lmcsheading{}{\pageref{LastPage}}{}{}%
{Dec.~09,~2021}{Jul.~13,~2023}{}


\begin{document}

\title[Impredicative Reduction Free Normalization]{Reduction Free Normalization for a\texorpdfstring{\\}{ }proof-irrelevant type of propositions}

\author[T. Coquand]{Thierry Coquand}
\address{Computer Science Department, University of Gothenburg}
\email{Thierry.Coquand@cse.gu.se}

\newcommand{\lift}[1]{\overline{#1}}
\newcommand{\ev}[1]{{\langle #1 \rangle}}
\newcommand{\Val}[1]{\ev{#1}}
\newcommand{\UU}{{\mathsf{U}}}
\newcommand{\VV}{\mathcal{V}}
\newcommand{\GU}{\mathcal{U}}
\newcommand{\GV}{\mathcal{V}}
\newcommand{\WW}{\mathsf{W}}
\newcommand{\Prop}{\mathsf{U_0}}
\newcommand{\Term}{\mathsf{Term}}
\newcommand{\mquote}{\mathsf{q}}
\newcommand{\unquote}{\mathsf{r}}
\newcommand{\El}{\mathsf{Elem}}
\newcommand{\Con}{\mathsf{Con}}
\newcommand{\Elem}{\mathsf{Elem}}
\newcommand{\id}{\mathsf{id}}
\newcommand{\pp}{\mathsf{p}}
\newcommand{\qq}{\mathsf{q}}
\newcommand{\fst}{\mathsf{fst}}
\newcommand{\snd}{\mathsf{snd}}
\newcommand{\mk}{\mathsf{mk}}
\newcommand{\REIFY}{\mathsf{reify}}
\newcommand{\NF}{\mathsf{nf}}
\newcommand{\CC}{{\mathcal C}}
\newcommand{\app}{\mathsf{app}}
\newcommand{\Neut}{\mathsf{Neut}}
\newcommand{\Norm}{\mathsf{Norm}}
\newcommand{\Var}{\mathsf{Var}}
\newcommand{\Type}{\mathsf{Type}}

\begin{abstract}
  \noindent We show normalization for a type theory with a hierarchy of universes and a proof irrelevant type of propositions, close
  to the type system used in the proof assistant Lean.
  The proof uses the technique of Artin glueing between the term model and a suitable preseaf model.
  This can also be seen as a proof relevant version of Tait's computability argument.
\end{abstract}

\maketitle

\section*{Introduction}

We show normalization and decidability of conversion
for dependent type theory 
with a cumulative sequence of universes $\UU_0,\UU_1\dots$  
with $\eta$-conversion \emph{and} where the type $\Prop$ is an impredicative
universe of proof-irrelevant propositions.
 One interest of such a system is that it is very close to the
type system used by the proof assistant Lean \cite{Carneiro19}.

Such a system with a hierarchy of universes, with the lowest level
impredicative, was introduced in \cite{coq86}. It was conjectured there
that this system is stronger than Zermelo set theory (without even introducing
primitive data types). This conjecture was solved by A. Miquel in \cite{Miquel04},
by encoding a non well-founded version of set theory where a set is interpreted
as a pointed graph up to bissimulation.
The notion of proof-irrelevant propositions goes back to de Bruijn \cite{AUT4}.

Our proof is a direct adaptation of the normalization argument
presented in \cite{coq:canon}. We recall three features of this approach
\begin{enumerate}
\item we never need to consider a \emph{reduction} relation,
\item we only define a reducibility \emph{predicate}, and this
  reducibility predicate is \emph{proof-relevant}\footnote{A key point
is to define reducibility as a \emph{structure} and not only
as a \emph{property}. It is only for the lowest impredicative universe
$\Prop$ that reducibility is a property.},
\item the reducibility predicate is not defined by an inductive-recursive
  relation.
\end{enumerate}

This approach has been much refined in \cite{Sterling22,Gratzer22}.
One goal of this note is to illustrate further the flexibility of this ``reduction free'' approach,
by combining it with an idea already used in \cite{AbelCP09} for dealing with proof irrelevance.
To each type $A$ in a context $\Gamma$,
we associate a set of syntactical expressions
$\Term(\Gamma,A)$ and a set $\Elem(\Gamma,A)$ of expressions \emph{modulo conversion}. We
have a quotient map $\Term(\Gamma,A)\rightarrow\Elem(\Gamma,A)$ and the main result
(Theorem \ref{mainth}) is to show that this map has a
section.

The metatheory used in the present note is the impredicative intuitionistic set theory 
IZFu$_{\omega}$,  introduced by P. Aczel \cite{aczel:relate}.
(Essentially the same argument works in a predicative version CZFu$_{\omega}$
for a predicative universe of proof-irrelevant propositions.)

As in the previous work \cite{coq:canon}, the approach is \emph{algebraic}.
We first define a general operation which associates to any model $M$ another 
\emph{normalization model} $M^*$ with a projection map $M^*\rightarrow M$.
We apply then this general construction to the initial model
to deduce various syntactical properties, such as normalization, decidability of conversion and type-checking.

\section{What is a model of type theory}

\subsection{Definition}

 We present a formal system, which at the same time can be thought of  describing  the
syntax of basic dependent type theory, with \emph{explicit substitutions} and
a \emph{name-free} (de Bruijn index) presentation, and defining what is a model
of type theory.

 A model of type theory consists of one set $\Con$ of \emph{contexts}. If
$\Gamma$ and $\Delta$ are in $\Con$ they determine a set $\Delta\rightarrow\Gamma$
of \emph{substitutions}. If $\Gamma$ is in $\Con$, it determines a set
$\Type(\Gamma)$ of \emph{types} in the context $\Gamma$. Finally, if
$\Gamma$ is in $\Con$ and $A$ is in $\Type(\Gamma)$ then this determines a set
$\Elem(\Gamma,A)$ of elements of type $A$ in the context $\Gamma$.

 This describes the \emph{sort} of type theory. We describe now the \emph{operations}
and the equations they have to satisfy. For any context $\Gamma$
we have an identity substitution $\id:\Gamma\rightarrow\Gamma$.
We also have a composition operator $\sigma\delta:\Theta\rightarrow\Gamma$ if
$\delta:\Theta\rightarrow\Delta$ and $\sigma:\Delta\rightarrow\Gamma$. 
The equations are
$$\sigma~ \id = \id~ \sigma = \sigma~~~~~~~~(\theta\sigma)\delta = \theta(\sigma\delta)$$

We have a terminal context $1$ and for, any context $\Gamma$, a map
$():\Gamma\rightarrow 1$. Furthermore, $\sigma = ()$ if $\sigma:\Gamma\rightarrow 1$.

 If $A$ in $\Type(\Gamma)$ and $\sigma:\Delta\rightarrow\Gamma$
 we should have $A\sigma$ in $\Type(\Delta)$.
Furthermore, we have
$$A~\id = A~~~~~~(A\sigma)\delta = A(\sigma\delta)$$
If $a$ in $\Elem(\Gamma,A)$ and $\sigma:\Delta\rightarrow\Gamma$
we should have $a\sigma$ in $\Elem(\Delta,A\sigma)$.
Furthermore 
$$a~\id = a~~~~~~(a\sigma)\delta = a(\sigma\delta)$$

 We have a \emph{context extension operation}: if $A$ in $\Type(\Gamma)$ we have
a new context $\Gamma.A$. There is a projection
$\pp:\Gamma.A\rightarrow \Gamma$ and a special element
$\qq$ in $\Elem(\Gamma.A,A\pp)$. If $\sigma: \Delta\rightarrow \Gamma$ and
$A$ in $\Type(\Gamma)$ and $a$ in $\Elem(\Delta,A\sigma)$
we have
an extension operation $(\sigma,a):\Delta\rightarrow \Gamma.A$.
We should have 
$$\pp (\sigma,a) = \sigma~~~~~~~~~\qq (\sigma,a) = a~~~~~~~~~
(\sigma,a)\delta = (\sigma\delta,a\delta)~~~~~~~~~~(\pp,\qq) = \id$$

 If $a$ in $\Elem(\Gamma,A)$ we write $[a]= (\id,a):\Gamma\rightarrow \Gamma.A$.
 Thus if $B$ in $\Type(\Gamma.A)$ and $a$ in $\Elem(\Gamma,A)$
 we have $B[a]$ in $\Type(\Gamma)$.
 If furthermore $b$ in $\Elem(\Gamma.A,B)$ we have $b[a]$ in $\Elem(\Gamma,B[a])$.

 If $\sigma:\Delta\rightarrow\Gamma$ and $A$ in $\Type(\Gamma)$
 we define $\sigma^+:\Delta.A\sigma\rightarrow\Gamma.A$ to be
 $(\sigma\pp,\qq)$.

 The extension operation can then be defined as $(\sigma,u) = [u]\sigma^+$.
 Thus instead of the extension operation, we could have chosen the operations
 $[u]$ and $\sigma^+$ as primitive, like in \cite{EhrhardThesis}. Our argument is independent of this choice
 of primitive operations.

 We suppose furthermore one operation $\Pi~A~B$ such that
$\Pi~A~B$ in $\Type(\Gamma)$ if $A$ in $\Type(\Gamma)$ and $B$ in $\Type(\Gamma.A)$.
We should have $(\Pi~A~B)\sigma = \Pi~(A\sigma)~(B\sigma^+)$.

We have an abstraction operation $\lambda b$ in $\Elem(\Gamma,\Pi~A~B)$
for $b$ in $\Elem(\Gamma.A,B)$
and an application operation $c~a$ in $\Elem(\Gamma,B[a])$
for $c$ in $\Elem(\Gamma,\Pi~A~B)$ and $a$ in $\Elem(\Gamma,A)$.
These operations should satisfy the equations
$$
{(\lambda b)}~{a} = b[a],~~~~~~c = \lambda (c\pp~\qq),~~~~~
(\lambda b)\sigma = \lambda (b\sigma^+),~~~~
({c}~{a})\sigma = {c\sigma}~{(a\sigma)}
$$

We assume each set $\Type(\Gamma)$ to be stratified in $\Type_0(\Gamma)\subseteq \Type_1(\Gamma)\subseteq\dots$.

Each subset $\Type_n(\Gamma)$ is closed by dependent product, and we have
$\UU_n$ in $\Type_{n+1}(\Gamma)$ such that $\Elem(\Gamma,\UU_n) = \Type_n(\Gamma)$.

\medskip

Finally we assume $\UU_0$ to be {impredicative} and types in $\UU_0$ to be
{proof-irrelevant}. \emph{Impredicativity} means that $\Pi~A~B$ is in $\Type_0(\Gamma)$ if
$B$ is in $\Type_0(\Gamma.A)$ where $A$ can be \emph{any} type, and \emph{proof-irrelevance}
means that
$a_0 = a_1:\Elem(\Gamma,A)$ whenever $A$ is in $\Type_0(\Gamma)$ and $a_0$ and $a_1$
are in $\Elem(\Gamma,A)$.

 We think of types in $\Type_0(\Gamma)$ as proof-irrelevant propositions.

\medskip

Note that, in an arbitrary model we may have some equality of the
form\footnote{This can even be the case \emph{a priori} in the term model, though it follows
  from our proof that this is \emph{not} the case.}
$\Pi~A~B = \UU_0$ and the operations, like product operations, don't need to be
injective.

\subsection{Examples of Models}

Like for equational theories, there is always the \emph{terminal} model
where all sorts are interpreted by a singleton.

\medskip

P. Aczel in \cite{aczel:relate} provides a model in
in a impredicative intuitionistic set theory 
IZFu$_{\omega}$, with intuitionistic versions of Grothendieck universes
$\VV_0,\VV_1,\dots,\VV_{\omega}$.

A context is interpreted as a set in $\VV_{\omega}$, and $\Type(\Gamma)$
is interpreted by $\Gamma\rightarrow\VV_{\omega}$.
The lowest universe $\UU_0$ is interpreted by the set of \emph{truth values} $\VV_0$: the
set of subsets of $1 = \{0\}$. In order to interpret the fact that $\UU_0$ is closed
by arbitrary products, P. Aczel introduces a non-standard encoding of dependent products,
see \cite{aczel:relate}, which we use in building our normalization model (see Appendix).
This encoding of dependent products $\Pi_{x\in A}B(x)$ is such that
$\Pi_{x\in A}B(x)\subseteq 1$ if we have $B(x)\subseteq 1$ for all $x$ in $A$.

\medskip

M. Hofmann \cite{Hofmann95} shows how to refine a presheaf model over an arbitrary
small category to a model of type theory. It models universes, and if
we use Aczel's encoding of dependent products, we also get a model where the
lowest universe $\GU_0$ is interpreted by the presheaf of sieves. Using Aczel's non-standard
encoding \cite{aczel:relate} of dependent products mentioned above, we see that
$\GU_0$ is closed by dependent products of families valued in $\GU_0$.
We write $\GU_0,~\GU_1,~\dots$ for
the universes corresponding to $\VV_0,~\VV_1,~\dots$

\medskip

We will work in the last section with the \emph{initial} or \emph{term} model $M_0$ (see Appendix). This is the model
where elements are syntactical expressions \emph{modulo}  equations/conversion rules.
One important result which follows from the ``normalization model'' we present in the next
section, is that equality is \emph{decidable} for the initial model, and that \emph{constructors
are injective}; this means in particular that we cannot have an equality of the form
$\UU_0 = \Pi~A~B$ and that $\Pi~A_0~B_0 = \Pi~A_1~B_1$ in $\Type(\Gamma)$ implies
$A_0 = A_1$ in $\Type(\Gamma)$ and $B_0 = B_1$ in $\Type(\Gamma.A_0)$. This injectivity
property may not hold in general for an arbitrary model; for instance in the set model, we have
$\emptyset^A = \emptyset$ for any non empty set $A$.

\section{Normalization Model}

We present a variation of the model used in \cite{coq:canon}.
As in \cite{coq:canon}, we work in a suitable \emph{presheaf} topos, but with a slight variation
for the choice of the base category. We start from an arbitrary model $M$.

\subsection{Category of telescopes} 

As in \cite{coq:canon}, we define first the collection of \emph{telescopes} $X,Y,Z,\dots$. These are finite
list $X = A_0,A_1,\dots ,A_{n-1}$ with $A_0$ in $\Type()$, $A_1$ in $\Type(().A_0)$ and so on.
Any telescope $X$ has an interpretation $\ev{X}$ which is a context of the model $M$, by taking
$\ev{X} = ().A_0.A_1.\dots.A_{n-1}$. We can have $\ev{X} = \ev{Y}$ in $M$ without having $X = Y$.
We write $()$ the empty telescope. If $X$ is a telescope and $A$ in $\Type\ev{X}$, we may write $X.A$
for $X,A$.

We can now define the base category of the presheaf model.
A map $\alpha:Y\rightarrow_S X$ is a syntactical object defined inductively. We have $():Y\rightarrow_S ()$
and if we have already define $\alpha:Y\rightarrow_S X$ then we can either add a type to $Y$
getting $\alpha\pp : Y,B\rightarrow_S X$, or we can add a type to $X$, getting
$\alpha^+ : Y,A\ev{\alpha}\rightarrow_S X,A$. We define at the same time $\ev{()}$ by the clauses:
$$\ev{()} = ()~~~~~~~~~\ev{\alpha\pp} = \ev{\alpha}\pp~~~~~~~~~\ev{\alpha^+} = \ev{\alpha}^+$$
A map $\alpha:Y\rightarrow_S X$ can be seen as a proof relevant witness that $Y$ extends $X$ (which was the
relation used in \cite{coqgal}). It is direct to define a syntactical identity map $\id_S : X\rightarrow_S X$
by induction on $X$ so that $\ev{\id_S} = \id$ and to define a composition operation. We get in this way
a category $\CC$ of telescopes\footnote{It would also have been possible to use \emph{renaming} as maps, as
in \cite{coq:canon,Sterling22,Gratzer22}.}.

We can also define a syntactic projection map $\pp_S:X.A\rightarrow X$ such that $\ev{\pp_S} = \pp$
by induction on $X$.
This category of syntactic extensions will be the base category
$\CC$ for the presheaf topos $\hat{\CC}$ in which we define the normalization
model\footnote{The use of context as world for a normalization argument goes back
to \cite{coqgal}. It was introduced there as a solution of the problem of having empty types, problem which was
solved in \cite{Girard71} by the introduction of a constant in all types with special reduction rules.}.

\subsection{Syntactic expressions} 

We introduce, for $A$ in $\Type\ev{X}$,
the set $\Term(X,A)$. This is a set of \emph{syntactical expressions}.
Contrary to the set $\Elem(\ev{X},A)$, these expressions are \emph{not}
quotiented up to conversion. Also the syntactical expressions don't use
explicit substitutions and can be thought of as annotated $\lambda$-expressions.

\medskip

The \emph{syntactical} expressions are described by the following grammar
$$
K,L,k~~::=~~v_n~|~\UU_n~|~\app~K~L~k~k~|~\lambda~K~K~k~|~\Pi~K~L~|~0
$$
where $v_n$ are de Bruijn index.
This forms a set with a \emph{decidable} equality. We define then inductively
for $A$ in $\Type\ev{X}$ a \emph{subset} $\Term(X,A)$ of this set
of syntactical expressions. Each such set $\Term(X,A)$ is then also a set
with a decidable equality. If $k$ is in $\Term(X,A)$ we define
by induction on $k$ an element $\ev{k}$ in $\Elem(\ev{X},A)$. This can be
thought of as the \emph{interpretation} of the syntactical expression $k$.
We can also see the map
$$
k\mapsto \ev{k}~~~~~~~~~~\Term(X,A)\rightarrow\Elem(\ev{X},A)$$
as a \emph{quotient} map.

\medskip

We have $\UU_n$ in $\Term(X,\UU_l)$ if $n<l$ and $\ev{\UU_n} = \UU_n$.

\medskip

We have $v_0$ in $\Term(X.A,A\pp)$
and $v_{n+1}$ in $\Term(X.A,B\pp)$ if $v_n$ is in $\Term(X,B)$.

We let $\ev{v_n}$ to be $\qq\pp^n$ and $\ev{\UU_l} = \UU_l$.

\medskip

If $K$ is in $\Term(X,\UU_n)$ and $L$ in $\Term(X.\ev{K},\UU_n)$
then $\Pi~K~L$ is in $\Term(X,\UU_n)$
and $\ev{\Pi~K~L} = \Pi~\ev{K}~\ev{L}$.
If furthermore $k'$ is in $\Term(X,\ev{\Pi~K~L})$ and
$k$ in $\Term(X,\ev{K})$ then
$\app~K~L~k'~k$ is in $\Term(X,\ev{L}[\ev{k}])$
and then $\ev{\app~K~L~k'~k} = \ev{k'}~\ev{k}$.

\medskip

If $K$ is in $\Term(X,\UU_n)$ and $L$ in $\Term(X.\ev{K},\UU_n)$
and $t$ is in $\Term(X.\ev{K},\ev{L})$
then $\lambda~K~L~t$ is in $\Term(X,\ev{\Pi~K~L})$
and $\ev{\lambda~K~L~t} = \lambda~\ev{t}$.

\medskip

If $K$ is in $\Term(X,\UU_l)$ and $l\leq n$ then $K$ is in $\Term(X,\UU_n)$.

\medskip

One \emph{key} addition to this notion of syntactical expressions,
introduced in order to deal with proof-irrelevant propositions,
is the special constant $0$. We have
$0$ in $\Term(X,A)$ whenever $A$ is in $\Type_0(\ev{X})$ and
$\Elem(\ev{X},A)$ is \emph{inhabited}.

Since $\Elem(\ev{X},A)$
is a \emph{subsingleton} we can define $\ev{0}$ to be any element $u$ of $\Elem(\ev{X},A)$.
(We don't need any choice since $u=v$ if $u$ and $v$ are in $\Elem(\ev{X},A)$.)

\medskip

If $u$ is in $\Elem(\ev{X},A)$ we write $\Term(X,A)|u$ the subset of
syntactical expressions $k$ in $\Term(X,A)$ such that $\ev{k} = u$.

\medskip

Like in \cite{coq:canon}, we need to define two subsets of $\Term(X,A)$, the subsets $\Norm(X,A)$ of \emph{normal}
terms and $\Neut(X,A)$ of \emph{neutral} terms. These are defined inductively by the following clauses.

\medskip

We have $v_0$ in $\Neut(X.A,A\pp)$
and $v_{n+1}$ in $\Neut(X.A,B\pp)$ if $v_n$ is in $\Neut(X,B)$.

\medskip

We have $\app~K~L~k~t$ in $\Neut(X,\ev{L}[\ev{t}])$ if $K$ in $\Norm(X,\UU_n)$ and $L$ in
$\Norm(X.\ev{K},\UU_n)$ and $k$ in $\Neut(X,\ev{\Pi~K~L})$ and $t$ in $\Norm(X,\ev{K})$.

\medskip

We have $\lambda~K~L~t$ in $\Norm(X,\ev{\Pi~K~L}[\ev{t}])$ if $K$ in $\Norm(X,\UU_n)$ and $L$ in\linebreak 
$\Norm(X.\ev{K},\UU_n)$ and $k$ in $\Neut(X,\ev{\Pi~K~L})$ and $t$ in $\Norm(X,\ev{K})$.

\medskip

We have $\Pi~K~L$ in $\Norm(X,\UU_n)$ if $K$ in $\Norm(X,\UU_n)$ and $L$ in
$\Norm(X.\ev{K},\UU_n)$.

\medskip

We have $K$ in $\Norm(X,\UU_n)$ if $K$ is in $\Neut(X,\UU_l)$ and $l\leq n$.

\medskip

We have $\UU_l$ in $\Norm(X,\UU_n)$ if $l<n$.

\medskip

We have $0$ in $\Norm(X,K)$ if $K$ is in $\Neut(X,\UU_0)$ and $\Elem(\ev{X},\ev{K})$ is inhabited

\medskip

We have $k$ in $\Norm(X,K)$ if $K$ is in $\Neut(X,\UU_n)$ with $n>0$ and $k$ is in $\Neut(X,K)$.

\medskip

As in \cite{Hofmann95,coq:canon}, we freely use the notations of type theory for
operations in the presheaf topos $\hat{\CC}$. In this presheaf models we have
a cumulative sequence of universe $\GU_n$, for $n = 0,1,\dots,\omega$.
Furthermore, as noticed above, $\GU_0$ inherits from $\VV_0$ the fact that it is closed by
arbitrary products.

In this model, we have a family of types $\Type_n$ (in the universe $\GU_1$) with families of types $\Elem(T)$ 
and $\Term(T)$ for $T:\Type_n$. We have two subtypes $\Norm(T)$ and $\Neut(T)$ of $\Term(T)$.
We also have an interpretation function $\Term(T)\rightarrow\Elem(T)$. Because of our choice
of morphisms for the category of telescopes, each $\Term(T)$ has (internally) a decidable equality.

\subsection{Artin Glueing} 

We define now a pseudomorphism \cite{KaposiHS19} between the model $M$ and the
presheaf model $\hat{\CC}$. The normalisation model $M^*$ will be a refinement of the glued
model \cite{KaposiHS19} along this pseudomorphism.

To each context $\Gamma$ in $M$, we associate a presheaf $|\Gamma|$ of $\hat{\CC}$ by taking
$|\Gamma|(X)$ to be the set $\ev{X}\rightarrow \Gamma$, with restriction
maps $\rho\mapsto \rho\alpha = \rho\ev{\alpha}$ for $\alpha:Y\rightarrow_S X$.

Each element $A$ in $\Type_n(\Gamma)$ in the model $M$ defines then
a presheaf map $|A|:|\Gamma|\rightarrow\Type_n$, by $\rho\mapsto A{\rho}$.
Similarly, each element $a$ in $\Elem(\Gamma,A)$ in the model $M$ defines a global
element $|a|:\Pi_{\rho:|\Gamma|}\Elem(|A|\rho)$.

For any $A$ in $\Type_n(\Gamma)$ in $M$, we have a constant in the presheaf model $\hat{\CC}$ 
$$\mk : \Pi_{\rho:|\Gamma|}|A|\rho\rightarrow |\Gamma.A|$$
and projections $\fst:|\Gamma.A|\rightarrow |\Gamma|$ and $\snd:\Pi_{\nu:|\Gamma.A|}|A|(\fst~\nu)$ satisfying the equations
$$
\fst~ (\mk~\rho~u) = \rho~~~~~~\snd~(\mk~\rho~u) = u~~~~~~~\nu = \mk~(\fst~\nu)~(\snd~\nu)
$$
This defines a pseudomorphism between the model $M$ and the model $\hat{\CC}$.


Given $B$ in $\Type_n(\Gamma.A)$, let us write $C = \Pi~A~B$ in $\Type_n(\Gamma)$.
If $\rho:|\Gamma|$ and $w$ in $\Elem(|C|\rho)$
and $u$ in $\Elem(|A|\rho)$ we can define $w~u$ in $\Elem(|B|(\mk~\rho~ u))$, which is levelwise the application.

\begin{lem}\label{syntax}
  In the presheaf topos $\hat{\CC}$, we have the following operations, for $\rho:|\Gamma|$ and
  $K : \Norm(\UU_n)$ such that $\ev{K} = |A|\rho$ and $G : \Pi_{k:\Neut(A\rho)}\Norm(\UU_n)$
  such that $\ev{G k} = |B|(\mk~ \rho ~\ev{k})$:
\begin{enumerate}
\item $\Pi_S~K~G : \Norm(\UU_n)$ such that $\ev{\Pi_S~K~G} = |C|\rho$,

\item $\lambda_S~g : \Norm(|C|\rho)|w$ for $w$ in $\Elem(|C|\rho)$ and
  $g : \Pi_{k:\Neut(A\rho)}\Norm(|B|(\mk~\rho~ \ev{k}))|(w~\ev{k})$,

\item $\app_S~K~G~k'~k : \Neut(|B|(\mk~\rho~ u))|(w~u)$ for $w$ in $\Elem(|C|\rho)$ and $u$ in $\Elem(|A|\rho)$
  and $k' : \Neut(|C|\rho)|w$ and $k : \Norm(|A|\rho)|u$.
\end{enumerate}
\end{lem}

\proof
  We prove the first point, the arguments for the two other points being similar.

  We have to define $\Pi_S~K~G$ in $\Term(X,\UU_n)$ such that
  $\ev{\Pi_S~K~G} = C{\rho}$. Here  $\rho$ is in $\ev{X}\rightarrow \Gamma$ and
  $K$ is in $\Norm(X,\UU_n)$ and such that $\ev{K} = A{\rho}$.
  Furthermore, $G$ is an operation such that
  $G{\alpha}~k$ is an element of $\Term(Y,\UU_n)$ satisfying
  $\ev{G{\alpha}~k}= B({\rho{\alpha}},\ev{k})$
  for ${\alpha}:Y\rightarrow X$ and $k$ in $\Term(Y,A\rho{\alpha})$
  and satisfying $(G{\alpha}~k)\alpha_1 = G({\alpha}\alpha_1)~(k\alpha_1)$,
  for $\alpha_1:Z\rightarrow_S Y$.

  We then take $\Pi_S~K~G$ to be $\Pi~K~(G\pp_S~v_0)$.

  We have $\ev{\Pi_S~K~G} = \Pi~\ev{K}~\ev{G\pp_S~v_0}$ and
  $\ev{K} = A\rho$ and $\ev{G\pp_S~v_0} = B(\rho\pp,\qq)$.

  If $\alpha:Y\rightarrow_S\ X$ we have
  $(\Pi_S~K~G)\alpha = (\Pi~K~(G\pp_S~v_0))\alpha = \Pi~K\alpha~(G\pp_S~v_0)(\alpha\pp_S,v_0)$
  and $(G\pp_S~v_0)(\alpha\pp_S,v_0) = G\alpha\pp_S~v_0$, so the operation $\Pi_S$ is functorial.
\qed

\section{Normalization model}

\subsection{Internal definitions}

 The first definitions are purely internal to the model $\hat{\CC}$.

For $T$ in $\Type_n$, we define $\Type_n'(T)$ to be the set of
4-tuples $(T',K,\mquote_T,\unquote_T)$ where\footnote{This definition
goes back to the unpublished paper \cite{AHS} for system F; one contribution of \cite{coq:canon} is to explain
how to treat universes and general dependent products, and the contribution of the present paper is to extend
this to an impredicative universe of proof irrelevant propositions.}
\begin{enumerate}
  \item $T'$ is in $\Elem(T)\rightarrow\GU_{n}$,
  \item $K$ is in $\Norm(\UU_n)|T$,
  \item $\mquote_T$, a ``quote'' function, is in $\Pi_{u:\Elem(T)}T'u\rightarrow\Norm(T)|u$,
  \item $\unquote_T$, a ``reflect'' function,  is in $\Pi_{k:\Neut(T)}T'\ev{k}$.
\end{enumerate}

\medskip

We define $\mquote_{\UU_n}~A~(A',K,\mquote_A,\unquote_A) = K$.

\medskip

 For $n>0$ and $K$ in $\Neut(\UU_n)$  we define
 $\unquote_{\UU_n}~K$ to be $(K',K,\mquote_K,\unquote_K)$
 where $K'u$ is $\Norm(K)|u$ and $\mquote_K~u~{u'} = u'$
 and $\unquote_K~k = k$.

 \medskip

 For $n = 0$, and $K$ in $\Neut(\UU_n)$, we define $\unquote_{\UU_n}~K$ to be $(K',K,\mquote_K,\unquote_K)$
 where $K'u$ is $\{0\}$ and\footnote{This is well-defined since $u$ is
   in $\Elem(\ev{K})$ and so $0$ is in $\Norm(\ev{K})$.}
 $\mquote_K~u~{u'} = 0$
 and $\unquote_K~k = 0$. 

 \subsection{The glued model for normalization}

We can now define the normalization model $M^*$, where
a context is a pair $\Gamma,\Gamma'$ where $\Gamma$ is a context of $M$
and $\Gamma'$ is a dependent family over $|\Gamma|$ in the model $\hat{\CC}$.

\medskip

We define $()'$ to be the constant family of constant presheaf $\{0\}$.

\medskip

 The set $\Type_n^*(\Gamma,\Gamma')$ is defined to be the set of pairs
$A,\lift{A}$ where $A$ is in $\Type_n^M(\Gamma)$ and
$\lift{A}$ is a global element of
$$\Pi_{\rho:|\Gamma|}\Gamma'(\rho)\rightarrow \Type_n'(|A|\rho)$$

\medskip

 An element of this type $A,\lift{A}$ is a pair $a,\lift{a}$ where $a$ is in $\Elem^M(\Gamma,A)$
 and $\lift{a}$ is a global element of
 $$\Pi_{\rho:|\Gamma|}\Pi_{\rho':\Gamma'(\rho)}\lift{A}\rho{\rho'}.1(|a|\rho)$$

We define $\lift{\UU_n} = \UU_n,{\Type'_n},\mquote_{\UU_n},\unquote_{\UU_n}$
and $\UU_n^*$ is the pair $\UU_n, \lift{\UU_n}$.

\medskip

The extension operation is defined by
$(\Gamma,\Gamma').(A,\lift{A}) = \Gamma.A,(\Gamma.A)'$
where $(\Gamma.A)'(\rho,u)$ is the set of pairs ${\rho'},{u'}$
with ${\rho'}\in \Gamma'(\rho)$ and ${u'}$ in $\lift{A}\rho {\rho'}.1(u)$.

\medskip

As in \cite{coq:canon}, we define a new operation
$\Pi^*~(A,\lift{A})~(B,\lift{B}) = C,\lift{C}$ where $C = \Pi~A~B$.

We write $(T',K,\mquote_T,\unquote_T) = \lift{A}\rho{\rho'}$ in $\Type_n'(|A|\rho)$
and for each $u$ in $\El(|A|\rho)$ and ${u'}$ in $T'(u)$ we write
$$(F'u {u'},F_0 u {u'},\mquote_F u {u'},\unquote_F u {u'})
   = \lift{B}(\mk~\rho~u)({\rho'},{u'})$$
   in  $\Type_n'(|B|(\mk~\rho~u))$. We then define $\lift{C}\rho{\rho'}$ in $\Type_n'(|C|\rho)$
   to be the tuple
\begin{itemize}
\item $R'(w) = \Pi_{u : \El(|A|\rho)} \Pi_{{u'} :  T'(u)}F' u {u'}({w}{u})$
\item $L = \Pi_S~ K ~G$ 
\item $\mquote_R~ w~{w'} = \lambda_S~ g$
\item $(\unquote_R~k) u {u'} = \unquote_F u {u'} (\app_S~K~G~{k}~(\mquote_T u {u'}))$  
\end{itemize}
where $G$ is the function $G~k  = F_0 \ev{k} (\unquote_T~k)$ and
$g$ the function $g~k = \mquote_F \ev{k} (\unquote_T k) ({w}~{\ev{k}})$ $({w'} \ev{k} (\unquote_Tk))$. 

\medskip

We can check using Lemma \ref{syntax}
that $R',L,\mquote_R,\unquote_R$ is an element of $\Type_n'(|C|\rho).$

\medskip

We get in this way a new model $M^*$ with a projection map $M^*\rightarrow M$.

\section{Applications of the normalization model}

For the term model $M_0$, we have an initial map $M_0\rightarrow M_0^*$
which is a section of this projection map. In this case, the contexts of $M_0$ are the same as telescopes
and we have $\ev{X} = X$.

\medskip

For each context $\Gamma$ of $M_0$, we can hence compute, using this section, $\Gamma'$ which is
\emph{internally} a dependent family over $|\Gamma|$. \emph{Externally},
this is given by a family of sets $\Gamma'(\Delta,\rho)$ for
$\rho:\Delta\rightarrow\Gamma$ with restriction maps
$\rho'\mapsto \rho'\alpha$ for $\alpha:\Delta_1\rightarrow_S \Delta$.

\medskip

For $A$ in $\Type(\Gamma)$ let us write
$A'\rho\rho'$ for $(\lift{A}\rho\rho').1$ and
$\unquote_A\rho\rho'$ for $(\lift{A}\rho\rho').4$, 
which, internally, is a function in $\Pi_{k:\Neut(|A|\rho)}A'\rho\rho'\ev{k}$,
Externally, this can be seen as a function
$\unquote_A(\Delta,\rho)\rho'k$ in $A'(\Delta,\rho)\rho'\ev{k}$
for $\rho:\Delta\rightarrow\Gamma$ and $\rho'$ in $\Gamma'(\Delta,\rho)$
and $k$ in $\Neut(\Delta,A\rho)$. This function satisfies
$(\unquote_A(\Delta,\rho)\rho'k)\alpha =
\unquote_A(\Delta_1,\rho\alpha)(\rho'\alpha)(k\alpha)$
for $\alpha:\Delta_1\rightarrow_S\Delta$.
Similarly we define $\mquote_A\rho\rho'$ to be $(\lift{A}\rho\rho').3$.

\medskip

For the two main applications of this normalization model,
we first build
${\id_{\Gamma}'}$ in $\Gamma'(\Gamma,\id)$. The definition is
by induction on $\Gamma$.

For $\Gamma = ()$ we take\footnote{We defined
  $\Gamma'(\Delta,\rho)$ to be the constant $1 = \{0\}$ in this case.}
${\id_{\Gamma}'} = 0$.

If we have defined ${\id_{\Gamma}'}$ in $\Gamma'(\Gamma,\id)$
and $A$ is in $\Type(\Gamma)$, let $\Delta = \Gamma.A$.
We have $\pp:\Delta\rightarrow\Gamma$ and $\pp_S:\Delta\rightarrow_S\Gamma$.
Let $\rho'$ be ${\id_{\Gamma}'}\pp_S$
in $\Gamma'(\Delta,\pp)$; we can define
${\id_{\Delta}'} = \rho', \unquote_A(\Delta,\pp)\rho' v_0$.

\medskip

If $A$ is in $\Type(\Gamma)$ we can compute
$\lift{A}~\id~{\id'} = (T',K,\mquote_T,\unquote_T)$
and we define $\REIFY(A)$ to be $(\lift{A}~\id~{\id'}).2 = K$.
We have $\ev{\REIFY(A)} = A$ since $\ev{\REIFY(A)} = A~\id = A$.
If furthermore $a$ is in $\Elem(\Gamma,A)$ we define
$\REIFY(a)$ in $\Norm(\Gamma,A)$ to be $\mquote_A~ \id ~\id' ~a~ (\lift{a}~\id~{\id'})$.
We have $\ev{\REIFY(a)} = a$ in $\Elem(\Gamma,A)$.

We can summarize this discussion as follows.

\begin{thm}\label{mainth}
  For each context $\Gamma$,
  the quotient map $k\mapsto \ev{k},~\Term(\Gamma,A)\rightarrow\Elem(\Gamma,A)$
  has   a section $a\mapsto \REIFY(a)$. 
\end{thm}

\begin{cor}
Equality in $M_0$ is decidable.
\end{cor}

\proof
If $a$ and $b$ are in $\Elem(\Gamma,A)$
we have $\REIFY(a) = \REIFY(b)$ in $\Term(\Gamma,A)$ if, and only if,
$a = b $ in $\Elem(\Gamma,A)$.
The result then
follows from the fact that the equality in $\Term(\Gamma,A)$ is decidable.
\qed

 We also can prove that $\Pi$ is one-to-one for conversions, following P. Hancock's argument
presented in \cite{ML75}.
The following Lemma follows from the definition of $\REIFY$.

\begin{lem} For $A$ in $Type(\Gamma)$ and $B$ in $\Type(\Gamma.A)$, we have 
  $\REIFY(\Pi~A~B) = \Pi~\REIFY(A)$ $\REIFY(B)$. 
\end{lem}

 \begin{cor}
   If $\Pi~A_0~B_0 = \Pi~A_1~B_1$ in $\Type(\Gamma)$ in the term model, we have
   $A_0=A_1$ in $\Type(\Gamma)$ and $B_0 = B_1$ in $\Type(\Gamma.A_0)$.
 \end{cor}
 \proof
   We have $\REIFY(\Pi~A_0~B_0) = \Pi~\REIFY(A_0)~\REIFY(B_0) = \Pi~\REIFY(A_1)~\REIFY(B_1) =$\linebreak$\REIFY(\Pi~A_1~B_1)$
   as syntactical expressions, and hence $\REIFY(A_0) = \REIFY(A_1)$. This implies
   $A_0 = A_1$ in $\Type(\Gamma)$.
   We then have $\REIFY(B_0) = \REIFY(B_1)$, which implies similarly
   $B_0 = B_1$ in $\Type(\Gamma.A_0)$.
 \qed
 
 \begin{cor}[Subject reduction]
    If $(\lambda b)~a$ is in $\Elem(\Gamma,D)$ then $b[a]$ is in $\Elem(\Gamma,D)$.
 \end{cor}
 
 \proof
   We have $b$ in $\Elem(\Gamma.A,B)$ and $a$ in $A'$ and $\lambda b$ in $\Pi~A'~B'$ with
   $\Pi~A~B = \Pi~A'~B'$ and $D = B'[a]$. By the previous Corollary, we have $A = A'$ and $B = B'$
   and $b[a]$ is in $\Elem(\Gamma,B[a]) = \Elem(\Gamma,B'[a]) = \Elem(\Gamma,D)$.
 \qed

 We can define a normal form function $\NF:\Term(\Gamma,A)\rightarrow\Norm(\Gamma,A)$
 by $\NF(k) = \REIFY(\ev{k})$.

 \medskip

 By mutual induction, we can show the following.

 \begin{lem}
   If $t$ is in $\Norm(\Gamma,A)$ then $t = \REIFY(\ev{t})$ and if $k$ is in $\Neut(\Gamma,A)$ then
   $\unquote_A~ \id ~\id' ~k = \lift{\ev{k}}~\id~\id'$.
 \end{lem}
 
 \begin{cor}
   We have $\NF(\NF(t)) = \NF(t)$ for any $t$ in $\Term(\Gamma,A)$.
 \end{cor}

 \begin{cor}
   The section map $\REIFY:\Elem(\Gamma,A)\rightarrow\Term(\Gamma,A)$ is natural in $\Gamma$ w.r.t. the
   morphisms in the telescope category $\CC$.
 \end{cor}

 \proof
   If $\alpha:\Delta\rightarrow\Gamma$ is a morphism in $\CC$ and $a$ is in
   $\Elem(\Gamma,A)$
   and $t = \REIFY(a)$ we have $t$ in $\Norm(\Gamma,A)$ and $t\alpha$ in $\Norm(\Delta,A\alpha)$
   with $\ev{t\alpha} = \ev{t}\alpha = a\alpha$. By the previous Lemma, we get
   $\REIFY(a\alpha) = \REIFY(\ev{t\alpha}) = t\alpha = \REIFY(a)\alpha$.
 \qed
 
 This implies that, in the presheaf model $\hat{\CC}$ the interpretation map $\Term(T)\rightarrow\Elem(T)$
 for $T$ in $\Type$ has a section $\Elem(T)\rightarrow \Term(T)$. Furthermore, $\Norm(T)$, which has
 internally a decidable equality, is isomorphic to $\Elem(T)$.

\section{Conclusion}

This note can be seen as a weak ``positive'' complement of the ``negative'' result
in \cite{AbelC20}, in the sense, that, in the absence of the problematic cast function
analysed in \cite{AbelC20}, we do have normalization and decidability of conversion.

 Our argument extends to the addition of dependent sum types with surjective
 pairing, or inductive types. In general, inductive types have to be declared
 in some universe $\UU_n$ with $n>0$.
 
 Note that it is possible to define the absurd proposition $\perp$ in $\UU_0$
 as $\Pi_{X:\UU_0}X$
 and to add the large elimination rule $\perp\rightarrow A$ for \emph{any} type $A$
 while preserving decidability of equality.

 A natural question is what happens if we consider a proof \emph{relevant} impredicative
 type of propositions. In a companion paper, we show that the present technique extends also to this case.
 
\section*{Acknowledgement}

Many thanks to Daniel Gratzer for many discussions on the topic of normalization. In particular, Daniel
convinced me that one could define the set of normal terms in such a way that one gets exactly the
$\beta$-normal \emph{$\eta$-long} normal terms (ideas also present in \cite{Sterling22}),
contrary to what I was doing in \cite{coq:canon}, and this was crucial for some results in the last section.
Many thanks also to the referee for many relevant comments.

\bibliographystyle{alphaurl}

\bibliography{nn}

\newpage

\appendix

 \section{Variations on the formulation of the system}

 Our formal system is not a generalised algebraic theory, presenting the sort $\Type(\Gamma)$ as stratified
 by $\Type_n(\Gamma)$ and requiring $\Elem(\Gamma,\UU_n) = \Type_n(\Gamma)$. It would instead have been possible 
 to use coercion functions $T_n(X)$ in $\Type(\Gamma)$ and
 $T_n^l(X)$ in $\Elem(\Gamma,\UU_l)$ for $X$ in $\Elem(\Gamma,\UU_n)$ with $T_l(T_n^l(X))  = T_n(X)$
 for $l\leq n$.
 One would then also need a dependent product operation $\Pi^n~X~Y$ in $\Elem(\Gamma,\UU_n)$
 with $T_n(\Pi^n~X~Y) = \Pi~T_n(X)~T_n(Y)$ for $n>0$ and the impredicative dependent product
 $\Pi^0~A~Y$ in $\Elem(\Gamma,\UU_0)$ with $T_0(\Pi^0~A~Y) = \Pi~A~T_0(Y)$.
 One can then apply e.g. \cite{PalmgrenV07} to justify the  existence of an initial model. We can see our
 system as an informal notation used to simplify the presentation.

One can wonder how crucial is the use of P. Aczel's encoding of dependent product \cite{aczel:relate}
which justifies the equality $T_0(\Pi^0~A~Y) = \Pi~A~T_0(Y)$. Without this encoding, we only have
one isomorphism between $T_0(\Pi^0~A~Y)$, which is a subset of $1$, and $\Pi~A~T_0(Y)$, which is also a
subsingleton, but may not be a subset of $1$. The following argument, due to M. Shulman, provides a more
modular solution to this issue, which is independent of the way one encodes dependent product in the
underlying set theory. One replaces the set model $M$ by a new model
$M'$ with the same notion of context but letting $\Type'(\Gamma)$ to be the disjoint sum
$\Type(\Gamma) + \Elem(\Gamma,\UU_0)$.
It is then possible to define by case a new product operation so that we get a strict equality
$T_0(\Pi^0~A~Y) = \Pi~A~T_0(Y)$.

\end{document}